\documentclass[a4paper ,12pt]{report}
\usepackage[left=0.7in,right=0.7in,top=0.7in,bottom=1.1in,includefoot,includehead,headheight=13.6pt]{geometry}

\usepackage{times}

\usepackage[T1]{fontenc}
\usepackage[english]{babel}
\usepackage{guit}
\title{\bf Risk Measures and Margining Control}
\date{}
\usepackage{amssymb}
\usepackage{amsmath}
\usepackage{eurosym}
\usepackage{makeidx}
\usepackage{mathrsfs}
\usepackage{eucal}
\usepackage{amscd}
\usepackage{mathtools}
\usepackage{booktabs}
\usepackage{upgreek}
\usepackage{amsthm}
\usepackage{multirow}

\usepackage{amsfonts}
\DeclarePairedDelimiter\ceil{\lceil}{\rceil}

\usepackage{booktabs}
\usepackage{svg}
\usepackage{listings}
\usepackage{xcolor}
\usepackage{mcode}
\usepackage{tikz}

\usepackage{bbm}
\usepackage{frontespizio}
\usepackage{fancyhdr}
\usepackage{subfig}
\usepackage{float}
 \usepackage{pgfplots}
\usepackage{graphicx}
\usepackage{epstopdf}
\usepackage{graphicx}
\usepackage{epstopdf}
\newcommand{\norm}[1]{\left\lVert#1\right\rVert}
\theoremstyle{definition}
\newtheorem{exmp}{Example}[section]
\newcommand{\vir}[1]{``#1''}

{\left\lbrace\begin{array}{@{}l@{}}}%
{\end{array}\right.}
\definecolor{lightgray}{gray}{0.5}
\author{Giuseppe Calafiore \\
Leonardo Massai}
\def\addlegendimage{\csname pgfplots@addlegendimage\endcsname}

\newcommand{\inv}{^{-1}}

\begin{document}

\pgfmathdeclarefunction{gauss}{3}{%
  \pgfmathparse{1/(#3*sqrt(2*pi))*exp(-((#1-#2)^2)/(2*#3^2))}%
}

\maketitle

\newpage
 \ 
 
\vspace{2cm}
{\em This document constitutes the final report of the contractual activity between Directa SIM and Dipartimento di Automatica e Informatica, Politecnico di Torino, on the research topic titled
``quantificazione del rischio di un portafoglio di strumenti finanziari per trading online su device fissi e mobili.''}

\vspace{1cm}
Torino, July 26, 2016

\newpage
\chapter{Introduction}
In this introductory chapter we are going to introduce some concepts and notations that will be used throughout this report.
\section{Simple and log-returns}
To begin with, we introduce the concept of simple and log return of a financial asset. Very generally speaking, a \emph{return} of an investment over a period of time is the ratio between the final value of the investment and its initial value.  Let $P_ {t-\Delta t}$ and $P_ {t}$ be the prices of an asset at the begining and at the end of a time period $\Delta t$ respectively. The \emph{simple return} of the asset over the period $\Delta t$ is given by:
 \begin{equation} 
RS_{\Delta t} :=  \frac{P_t}{P_ {t-\Delta t}}
  \end{equation} 
The \emph{rate of return}, instead, is a profit on an investment over a period of time, expressed as a proportion of the original investment:
 \begin{equation} 
r_{\Delta t} := \frac{P_t-P_ {t-\Delta t}}{P_ {t-\Delta t}}= \frac{P_t}{P_ {t-\Delta t}}-1
  \end{equation} 
  Notice that we have:
 \begin{equation} 
RS_{\Delta t}=1+ r_{\Delta t}
  \end{equation} 
Benefit of using returns, instead of prices, to describe market dynamics is normalization: measuring all variables in a comparable metric, thus enabling evaluation of analytic relationships among two or more variables despite originating from price series of unequal values.

In finance it is also useful to define  the so called \emph{log-return} as:
 \begin{equation} 
R_{\Delta t} :=  \ln {\left(\frac{P_t}{P_ {t-\Delta t}}\right)}=\ln{(1+r_{\Delta t})}
  \end{equation} 
 Log returns are of course different in value from simple returns, but a very useful approximation holds if simple returns are small enough. More precisely, if $r_{\Delta t} \to 0$, the following approximation holds\footnote{This comes from the limit $\lim_{x \to 0} \frac{\ln{(1+x)}}{x} = 1$}:
  \begin{equation} 
\ln{(r_{\Delta t}+1)} \approx r_{\Delta t}
  \end{equation}
  We may notice that returns are close to 0 for trades with short holding durations, i.e. when $\Delta t \to 0$.
  
  \medskip
 
Why should we use log-returns? there are a number of reasons to do that. The first and probably most important one one can be seen when we want to evaluate compounded returns. If we consider a sequence of $n$ investments or trades during time, then the total or compounded return, which is the running return of this sequence of trades over time, can be found by means of the simple returns for each of the $n$ trades as:
  \begin{equation}
\begin{split}
RS_n=\frac{P_n}{P_0} & = \frac{P_1}{P_0} \cdot \frac{P_2}{P_1}\dots  \frac{P_n}{P_{n-1}} \\
& = (1+r_1) (1+r_2) \dots (1+r_n) \\
& = \prod_{k=1}^n RS_k
\end{split}
\end{equation}
 Let us calculate the associated log-return:
  \begin{equation}
\begin{split}
R_n &=\ln{(RS_n)} \\
& = \ln{ ((1+r_1) (1+r_2) \dots (1+r_n))} \\
& = \ln{ (1+r_1)}+ \ln{ (1+r_2)} + \dots +  \ln{ (1+r_n)} \\
&= \sum_{k=1}^n  R_k
\end{split}
\end{equation}
If we compare formulae (1.6) and (1.7), we notice that, if we want to evaluate a simple compounded return, we need to calculate the product of many simple returns; instead, using log-returns we just need to sum other log-returns (this property is known as \emph{time-additivity} of log-returns). This is a dramatic semplification when, modeling the market, we consider returns as random variables. Indeed, a common assumption in finance is to consider log-returns as independent identically distributed  (i.i.d.) normal random variables. Because of formula (1.7), compounded log-returns are simply a sum of such variables, and it is a very well known fact that a finite sum of i.i.d. normal variables is also normally distributed. Further, this sum is useful even for cases in which returns diverge from normal, as the central limit theorem reminds us that the sample average of this sum will converge to normality (presuming finite first and second moments). Notice that nothing of this would be true if we used simple returns, since in general there are no simple results about the distribution of a product of random variables. 

Log-returns also prevent an issue regarding numerical stability: addition of small numbers is numerically safe, while multiplying small numbers is not as it is subject to arithmetic underflow. For many interesting problems, this is a serious potential problem. To solve this, either the algorithm must be modified to be numerically robust or it can be transformed into a numerically safe summation via logs.

\chapter{Risk Measures}

Risk is a key element in finance and a major activity in many financial institutions is to recognize the sources of risk, then manage and control them. This is only possible if risk is quantified and, as we will see, there are several ways to do that. The aim of this chapter is to introduce some risk measures used to model risk and to describe difference, pros and cons of each of these. The chapter first defines axiomatically what a risk measure and a deviation risk measure are by stating the properties that such a measures must have. Afterwards, some widely used risk measures are introduced.

\section{Axiomatic definition and coherent risk measures}
A risk measure is a function that tries to quantify the downside risk and it is used to determine the amount of an asset or set of assets (traditionally currency) to be kept in reserve. The purpose of this reserve is to make the risks taken by financial institutions, such as banks and insurance companies, acceptable to the regulator.
From a formal point of view, a risk measure is defined as a mapping from a set of random variables to the real numbers. In finance, this set of random variables represents portfolio returns. In other words: we are seeking for a function that takes as input all the data describing our portfolio (random returns and weights for each asset) and gives us a number describing the \emph{riskiness} of the portfolio as output. In 1999 Artzner et al. proposed a list of properties that any good risk measure should have and this list gave rise
to the concept of coherent and incoherent measures of risk. Since then a substantial body of research has
developed on the theoretical properties of risk measures and we describe some of these results here. 

\medskip

The common notation for a risk measure associated with a random variable $X$ is $\rho(X)$. Let $\mathcal{L}$ be the set of random variables describing the portfolio losses over a fixed time frame.
A risk measure is a map \begin{equation} \rho: \mathcal{L} \to \mathbb{R} \cup \{+\infty\} \end{equation}  that should have certain properties:

\begin{itemize}
\item \textbf{Normalized}

    $\rho(0) = 0$
    
    This is just a conventional value.
    
\item \textbf{Translative}
    
    $\mathrm{If}\; a \in \mathbb{R} \; \mathrm{and} \; Z \in \mathcal{L} ,\;\mathrm{then}\; \rho(Z + a) = \rho(Z)+a$
    
    This property states that if we add a costant and certain payoff/loss to our portfolio, that should be algebrically added from the risk. Indeed, if $a>0$ than we have a certain loss and we should increase the safety reserve to be kept. On the other hand, if $a<0$ the reserve should be reduced as we have a certain additional payoff available.
    
    \item \textbf{Monotone}
    
    $\mathrm{If}\; Z_1,Z_2 \in \mathcal{L} \;\mathrm{and}\; Z_1 \leq Z_2 ,\; \mathrm{almost \: surely, then} \; \rho(Z_1) \leq \rho(Z_2)$
    
    This last property is self-explaining and simply states that if a certain loss is almost surely smaller than another one, than the risk (the reserve to be kept) associated with the smaller loss should be smaller than the risk associated with the greater one.

\end{itemize}

In additional to the properties listed above,  theoreticians have added two more properties that define a so called \emph{coherent risk measure}:

\begin{itemize}
\item \textbf{Sub-additivity}

    $\mathrm{If}\; Z_1,Z_2 \in \mathcal{L} ,\; \mathrm{then}\; \rho(Z_1 + Z_2) \leq \rho(Z_1) + \rho(Z_2) $
    
Indeed, the risk of two portfolios together cannot get any worse than adding the two risks separately: this is the diversification principle.
    
\item \textbf{Positive homogeneity}
    
    $\mathrm{If}\; \alpha \ge 0 \; \mathrm{and} \; Z \in \mathcal{L} ,\; \mathrm{then} \; \rho(\alpha Z) = \alpha \rho(Z)$

Loosely speaking, if we double the investment in our portfolio then we double the risk.
\end{itemize}

Notice that these two additional properties imply that a coherent risk measure must be a convex functional. Indeed, if $\lambda \in [0,1]$ then we have:
 \begin{equation} \rho(\lambda X+(1-\lambda)Y) \le \rho(\lambda X)+\rho((1-\lambda)Y)=\lambda \rho(X)+(1-\lambda) \rho(Y) \end{equation} 
If $X$ and $Y$ describe random returns, then the random quantity $\lambda X+(1-\lambda)Y$ stands for the return of a fully invested portfolio composed of two financial instruments having returns $X$ and $Y$ and weights $\lambda$ and $1-\lambda$ respectively. Therefore, the convexity property states that the risk of a portfolio is not greater than the sum of the risks of its components, meaning that is is the convexity property which is behind the diversification effect that we expect in finance.
\newpage

\section{Value at risk}

A risk measure which has been widely accepted since 1990s is \emph{value at risk}. In the late 1980s, it was integrated by JP Morgan on a firmwide level into is risk-management system known as RiskMetrics. In the mid 1990s, the VaR measure was approved by regulators as a valid approach to evaluate capital reserves needed to cover market risk. 

\medskip

The idea behind the VaR is  rather simple: we try to evaluate the maximum loss that our portfolio could suffer with a given probability over a certain time horizon. In order to define VaR, we need infact two parameters: a given time horizon for which the loss must be calculated and a confidence level $\alpha \in [0,1]$. For a given portfolio, the VaR at confidence level $\alpha$, denoted in the following with $\mbox{VaR}_{\alpha}$, is defined as a threshold loss value, such that the probability that the loss on the portfolio over the given time horizon exceeds this value is exactly $\alpha$. Informally, VaR allows to say something like:
“We are $\alpha$ percent sure that we will not lose more than VaR dollars over the given time horizon”. For example, if a portfolio of stocks has a one-day 5\% VaR of 1 million, there is a 0.05 probability that the portfolio will fall in value by more than 1 million over a one day period if there is no trading. Informally, a loss of 1 million or more on this portfolio is expected on 1 day out of 20 days (because of 5\% probability). A loss which exceeds the VaR threshold is termed a "VaR break."

\medskip

The above description of VaR can be stated mathematically in a precise way:
let $L$ be the random variable describing the loss on the given portfolio over a certain time interval (one day, one month $\dots$), then the $\mbox{VaR}_{\alpha}$ is defined as follows:
\begin{equation}
\mbox{VaR}_{\alpha}(L):= \inf_l \{l \in \mathbb{R} \: : \: \mathbb{P}(L>l)\le \alpha\}=\inf_l \{ l \in \mathbb{R} \: : \: F_L(l) \ge 1-\alpha\}
\end{equation}
where $F_L$ is the loss cumulative distribution function. We see that VaR is actually the quantile function of the distribution of losses:
\[
\mbox{VaR}_{\alpha}(L)  = F_L\inv (1-\alpha) \doteq \inf_l \{ l \in \mathbb{R} \: : \: F_L(l) \ge 1-\alpha\},\quad \alpha\in(0,1)
\]
($F_L\inv$ is the regular inverse of $F_L$ if this function is strictly increasing).
Figure~\ref{fig:var:pdf} and Figure 2.1 show a graphical interpretation of VaR.

\begin{figure}[h]
\centering
\begin{tikzpicture}
% define normal distribution function 'normaltwo'
  \def\normaltwo{\x,{3*1/exp(((\x-4.5)^2)/5)}}
 
% input y parameter
    \def\y{6.3}
 
% this line calculates f(y)
    \def\fy{3*1/exp(((\y-4.5)^2)/5)}
% Shade orange area underneath curve.
    \fill [fill=orange!60] (6.3,0) -- plot[domain=6.3:9] (\normaltwo) -- ({\y},0) -- cycle;
 
% Draw and label normal distribution function
    \draw[color=blue,domain=0:9,very thick,cyan!50!black, smooth] plot (\normaltwo) node[right] {};
 
% Add dashed line dropping down from normal.
    \draw[dashed] ({\y},{\fy}) -- ({\y},0) node[below] {$\mbox{VaR}_{\alpha}$};

% Optional: Add axes
    \draw[thick, ->] (0,0) -- (9.5,0) node[right] {Loss};
    \node at (7, 0.4) {$\alpha$};
    
%    \node[draw=black,thick,rounded corners=2pt, smooth, below left=2mm] at (10.7,3.1) {%
%\begin{tabular}{@{}r@{ }l@{}}
% \raisebox{2pt}{\tikz{ \draw[color=blue,domain=0:9,very thick,cyan!50!black] (0,0) -- (5mm,0);}}&Loss density\\
% \raisebox{2pt}{\tikz{ \draw[color=blue,domain=0:9,very thick,cyan!50!yellow] (0,0) -- (5mm,0);}}&Loss cumulative
%\end{tabular}};
\end{tikzpicture}
\caption{Probability density function of the loss $L$.}
\label{fig:var:pdf}
\end{figure}
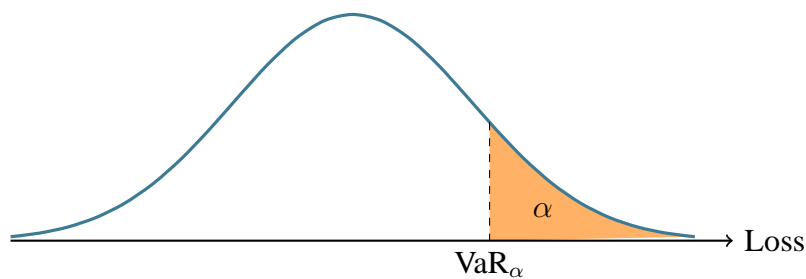

\begin{figure}[h]
\centering
\begin{tikzpicture}
% define normal distribution function 'normaltwo'
\def\normaltwo{\x,{1.9/(1+exp(-1.4*(\x-4.4))}}
 
% input y parameter
    \def\y{6.3}
 
% this line calculates f(y)
    \def\fy{1.9/(1+exp(-1.4*(\y-4.4))}
% Shade orange area underneath curve.

% Draw and label normal distribution function
    \draw[color=blue,domain=0:9,very thick,cyan!50!yellow, smooth] plot (\normaltwo) node[right] {};
 
% Add dashed line dropping down from normal.
    \draw[dashed] ({\y},{\fy}) -- ({\y},0) node[below] {$\mbox{VaR}_{\alpha}$};
    \draw[dashed] ({\y},{\fy}) -- (0,{\fy})  node[above right] {$1-\alpha$};
    
% Optional: Add axes
    \draw[thick, ->] (0,0) -- (9.5,0) node[right] {Loss};
        \draw[thick, ->] (0,0) -- (0,2.6) node[right] {};
\end{tikzpicture}
\caption{Cumulative distribution function of the loss $L$.}
\label{fig:var:pdf}
\end{figure}
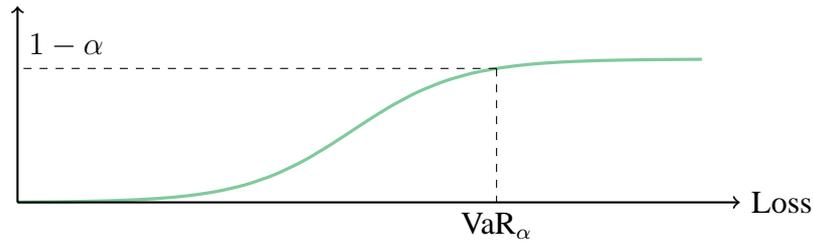

VaR can be equivalently expressed in terms of the random payoff $X = -L$ as follows:
\begin{equation}
\mbox{VaR}_{\alpha}(X)= -\sup_x \{x \in \mathbb{R} \: : \: \mathbb{P}(X < x)\le \alpha\}.
\label{eq:var:X}
\end{equation}
We observe that it holds that 
\[
\mbox{VaR}_{\alpha}(X) = \mbox{VaR}_{\alpha}(L) = F_L\inv (1-\alpha).
\]
Further, if the distribution of $X$ is continuous, then it also holds that 
\begin{equation}
\mbox{VaR}_{\alpha}(X)  =- F_X\inv(\alpha).
\label{eq:var:ret:cont}
\end{equation}
In the following we show that value at risk is actually a risk measure but not a coherent one. We assume that the random variable $L$ represents a random loss and we consider the definition (2.3). 

The normalized property is immediate to check so we focus on the other ones. 
We can use the definition to check the translative property; let $c \in \mathbb{R}$, then:
\begin{equation}
\begin{split}
\mbox{VaR}_{\alpha}(L+c) & = \inf_l \{l \in \mathbb{R} \: : \: \mathbb{P}(L+c>l)\le \alpha\} \\
& =\inf_l \{l \in \mathbb{R} \: : \: \mathbb{P}(L>l-c)\le \alpha\} \\
& = \inf_s \{s \in \mathbb{R} \: : \: \mathbb{P}(L>s)\le \alpha\}+c \\
&= \mbox{VaR}_{\alpha}(L)+c
\end{split}
\end{equation}
Notice that this property modifies a little bit if we consider the VaR of a return or a payoff. Indeed, if we use definition (2.4) we find:
\begin{equation}
\begin{split}
\mbox{VaR}_{\alpha}(X+c) & = -\sup_x \{x \in \mathbb{R} \: : \: \mathbb{P}(X+c < x)\le \alpha\} \\
& =-\sup_x \{x \in \mathbb{R} \: : \: \mathbb{P}(X < x-c)\leq \alpha\} \\
& = -\sup_s \{s \in \mathbb{R} \: : \: \mathbb{P}(X < s)\leq  \alpha\}-c \\
&= \mbox{VaR}_{\alpha}(X)-c
\end{split}
\end{equation}

We now check the monotonicity property; let  $L_1,L_2 \in \mathcal{L}  \;\mbox{be  two\:  random variable and}\; L_1 \leq L_2 ,\; \mbox{almost surely, then}$:
\begin{equation}
\begin{split}
\mbox{VaR}_{\alpha}(L_1) & = \inf_l \{l \in \mathbb{R} \: : \: \mathbb{P}(L_1>l)\le \alpha\} \\
& \le \inf_l \{l \in \mathbb{R} \: : \: \mathbb{P}(L_2>l)\le \alpha\} \\
& = \mbox{VaR}_{\alpha}(L_2)
\end{split}
\end{equation}

Since the three fundamental properties hold, then we proved that VaR is a risk measure. However, as we have already anticipated, it is not a coherent risk measure as in general it fails to satisfy the sub-additivity property. Indeed, there exist cases for which the VaR of a combined portfolio may be greater than the sum of the single VaR of its constituents, which clearly violates the diversification principle. 

We give an example of this. 
\begin{exmp}
Consider two zero-coupon bonds, whose issuer may default with a probability of 4\%, in which case we lose 100 (the face value of the bond).
The losses for the two bonds are two random variables ($X$ and $Y$, respectively) with the following probability mass function:
\[
X,Y=
\begin{cases}
0 & \text{with probability 96\%} \\
100 & \text{with probability 4\%} 
\end{cases}
\]
Let us now evaluate the 95\% confidence level-VaR. Since the probability of having no losses (96\%) for each bond is greater than the confidence level, we have:
\begin{equation}
\mbox{VaR}(X)=\mbox{VaR}(Y)=0
\end{equation}
What happens if we combine the two bonds in a portfolio? is is easy to evaluate the new distribution of losses. Indeed, the probability mass function of $X+Y$ is given by:
\[
X+Y=
\begin{cases}
0 & \text{with probability} \; 0.96^2=0.9216 \\
100 & \text{with probability} \; 2 \times 0.96 \times 0.04 = 0.0768  \\
200 & \text{with probability} \; 0.04^2=0.016
\end{cases}
\]
and in this case we see that $\mbox{VaR}(X+Y)=100>\mbox{VaR}(X)+\mbox{VaR}(Y)=0$.
\end{exmp}
Under some particular hypothesis, for example when the profit-loss distribution is one of the elliptical distribution family and has finite variance, then value at risk is sub-additive. However, being incapable, in general, of recognizing the diversification effect is one of the main drawbacks of VaR.

  Another important disadvantage of VaR is that it is not informative about the magnitude of the losses larger than VaR level. Because of its definition, there are cases for which two loss distributions have the same VaR at a certain confidence level, even though in one case the maximum potential loss is bounded while in the other case it is not. The next picture shows this phenomenom.
  \begin{figure}[h]
\centering
\begin{tikzpicture}
% define normal distribution function 'normaltwo'
  \def\normaltwo{\x,{2.5*1/exp(((\x-3.5)^2)/2.2)}}
 
% input y parameter
    \def\y{5}
 
% this line calculates f(y)
    \def\fy{2.5*1/exp(((\y-3.5)^2)/2.2)}
% Shade orange area underneath curve.
    \fill [fill=orange!60] (\y ,0) -- plot[domain=\y:6.5] (\normaltwo) -- ({\y},0) -- cycle;
 
% Draw and label normal distribution function
    \draw[color=blue,domain=0:6.5,very thick,cyan!50!black, smooth] plot (\normaltwo) node[right] {};
 
% Add dashed line dropping down from normal.
    \draw[thick, dashed] ({\y},{\fy}) -- ({\y},0) node[below] {$\mbox{VaR}$};

% Optional: Add axes
    \draw[thick, ->] (0,0) -- (6.8,0) node[right] {Loss};
 \node at (5.4, 0.22) {5\%};
   
\end{tikzpicture}
\begin{tikzpicture}
% define normal distribution function 'normaltwo'
  \def\normaltwo{\x,{2.5*1/exp(((\x-3.5)^2)/2.2)}}
 
% input y parameter
    \def\y{5}
 
% this line calculates f(y)
    \def\fy{2.5*1/exp(((\y-3.5)^2)/2.2)}
% Shade orange area underneath curve.
    \fill [fill=orange!60]({\y},0.002) rectangle (5.7, {\fy});
 
% Draw and label normal distribution function
    \draw[color=blue,domain=0:5,very thick,cyan!50!black, smooth] plot (\normaltwo) node[right] {};
 
     \draw[color=blue,domain=0:6,very thick,cyan!50!black, smooth]({\y},{\fy}) -- (5.7,{\fy})node[right] {};
      \draw[color=blue,domain=0:6,very thick,cyan!50!black, smooth](5.7,0) -- (5.7,{\fy})node[right] {};
% Add dashed line dropping down from normal.
    \draw[thick, dashed] ({\y},{\fy}) -- ({\y},0) node[below] {$\mbox{VaR}$};

% Optional: Add axes
    \draw[thick, ->] (0,0) -- (6.5,0) node[right] {Loss};
    \node at (5.4, 0.38) {5\%};
   
\end{tikzpicture}
\caption{Unbounded versus bounded maximum loss with same VaR at 5\%.}
\end{figure}
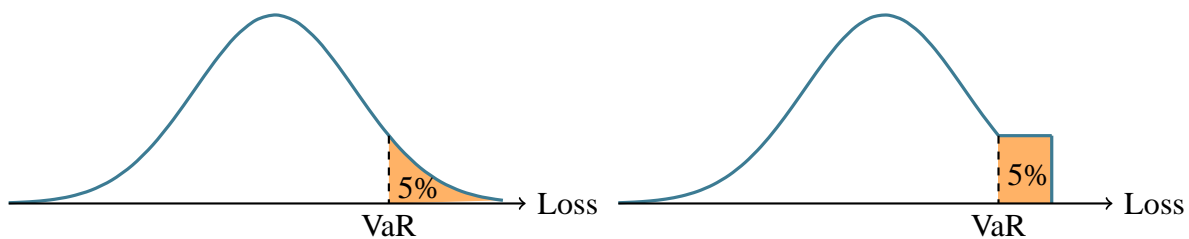

It is clear that we should assign a different risk to these situations, as in one case we cannot lose more than a certain amount of money while in the other case the loss is potentially infinite. VaR is not able to tell us much about the \vir{bad} tail of the  loss/profit distribution and it may underestimate extreme events. 

\subsection{Estimating Value at risk}
One of the main reason why VaR is so widely used is because it is rather simple to compute in practice. In the following we will discuss three approaches for portfolio VaR calculation which are used in practice. 

\subsubsection{The approach of RiskMetrics}
The approach of RiskMetrics Group is centered on the assumption that stock returns (continuously compounded) have a multivariate normal distribution. Under this assumption, the distribution of the portfolio, which is a weighted sum of the stock returns, is also normal. Therefore, in order to calculate the portfolio VaR, we only need to calculate the expected value and the variance of the portfolio return, which is an easy task because of the normality assumption. Notice that the assumption of normally distributed returns is consistent with the Black-Scholes model of option pricing, for which stock prices folow a lognormal distribution.

In the following we describe the method more formally in its simplest form. We consider a portfolio consisting of $n$ assets, whose random daily (rate of) returns are denoted with $X_1,X_2, \dots, X_n$; the portfolio weights associated to them are $w_1,w_2,\dots,w_n$. Thus, the portfolio daily return can be calculated as:
\begin{equation}
R_p=w_1 X_1+w_2 X_2+\dots+w_n X_n 
\end{equation}
It is immediate to find the expected daily return and daily variance of $R_p$; let us start we the expected value:
\begin{equation}
\mathbb{E}[R_p]=w_1 \mathbb{E}[X_1]+w_2 \mathbb{E}[X_2]+\dots+w_n \mathbb{E}[X_n]= \sum_{k=1}^{n} w_k \mathbb{E}[X_k]
\end{equation}
The daily variance is given by:
\begin{equation}
\sigma_{R_p}^2=w_1^2 \sigma_{X_1}^2+w_2^2 \sigma_{X_2}^2+\dots+w_n^2 \sigma_{X_n}^2+\sum_{i \ne j} w_i w_j \mathrm{Cov}(X_i, X_j)
\end{equation}
Notice that, since  $R_p \sim \mathcal{N}(\mu_R, \sigma_{R_p}^2)$, then, by normalizing, we have that
\begin{equation}
z\doteq \frac{R_p-\mu_R}{\sigma_{R_p}} \sim \mathcal{N}(0,1),
\end{equation}
or, equivalently,
\[
R_p = \sigma_{R_p} z + \mu_R, \quad z\sim \mathcal{N}(0,1).
\]
Then, 
\begin{equation}
F_{R_p} (x) = \mathbb{P}(R_p \le x) = 
\mathbb{P}\left( z
\le \frac{x-\mu_R}{\sigma_{R_p}} \right) = 
\Phi\left(\frac{x-\mu_R}{\sigma_{R_p}} \right),
\end{equation}
where $ \Phi$ is the standard Normal cumulative function.

Now, since the distribution is continuous, by using (\ref{eq:var:ret:cont}), we obtain that the daily $\mbox{VaR}_{\alpha}$ of the portfolio return is 
\begin{equation}
\begin{split}
\mbox{VaR}_{\alpha}(R_p) & = -\sup_x \{ x \in \mathbb{R} \: : \: \mathbb{P}(R_p  < x)\le \alpha\} \\
& = -\sup_x \left\{ x \in \mathbb{R} \: : \: \mathbb{P}\left(z < \frac{x-\mu_R}{\sigma_{R_p}} \right)\le \alpha\ \right\} \\
\mbox{[since $z$ is Normal]} & =- \sup_x \left\{ x \in \mathbb{R} \: : \: 
\Phi \left(  \frac{x-\mu_R}{\sigma_{R_p}} \right) \leq \alpha \right\} \\
 & = -\sup_x \left\{ x \in \mathbb{R} \: : \: 
 \frac{x-\mu_R}{\sigma_{R_p}} \leq \Phi\inv(\alpha) \right\} \\
& =  -\Phi^{-1}(\alpha) \sigma_{R_p} -\mu_R \\
&= \Phi^{-1}(1-\alpha) \sigma_{R_p} -\mu_R
\end{split}
\end{equation}

Equivalently, we can obtain the same result by using the properties of VaR as risk measure:
\begin{equation}
\mbox{VaR}_{\alpha}\left(\frac{R_p-\mu_{R_p}}{\sigma_{R_p}} \right)  = \frac{1}{\sigma_{R_p}} \mbox{VaR}_{\alpha}(R_p-\mu_{R_p}) = \frac{1}{\sigma_{R_p}} (\mbox{VaR}_{\alpha} (R_p)+\mu_{R_p})
\end{equation}
and expliciting $\mbox{VaR}(R_p)$, we get:
\begin{equation}
\mbox{VaR}_{\alpha} (R_p)= \mbox{VaR}_{\alpha}\left(\frac{R_p-\mu_R}{\sigma_{R_p}} \right) \sigma_{R_p}-\mu_R
\end{equation}
we can notice that $\mbox{VaR}_{\alpha}\left(\frac{R_p-\mu_R}{\sigma_{R_p}} \right) $ is the $\alpha$-level VaR of a standard normal variable;  by definition this is just the $(1-\alpha)$-quantile of a standard normal and the result follows immediately.

\medskip

So far we have shown how to calculate the daily VaR starting from the knowledge of daily returns and variances. What if we want to evaluate the VaR over a different time horizon? In principle, we should know the portfolio return distribution for the chosen time frame. The RiskMetrics approach allows us to do that with a simple method, known as the \emph{square root of time rule}, which essentially says that the standart deviation of returns scales with the square root of the time period. In order to explain the concept, we consider a simplified scenario, in which we can describe daily continuously compounded returns by means of i.i.d. random variables $X_1, X_2, \dots, X_T$ with mean value $\mu$ and variance $\sigma^2$; here $T$ defines the time interval. Let us evaluate mean return and standard deviation over the interval. The total return over the interval can be expressed as $Y=\sum_{k=1}^T X_k$. Then we have:
\begin{align}
 \mathbb{E}[Y] &= \sum_{k=1}^T \mathbb{E}[X_k]=\mu T  \\[2ex]
\sigma_Y & = \sqrt{\sum_{k=1}^T \mathrm{Var}(X_k)}=\sigma \sqrt{T} \notag 
\end{align}
We see that the expected value scales linearly with time, whereas the standard
deviation scales with the square root of time. Here the square root of time rule has been obtained considering i.i.d. variables, which is a strong and not realistic hypothesis, but it can be shown to hold true even in more general situations. In the case of the RiskMetrics approach for example, variances are considered time variant but the rule still applies. The square root of time rule is a consequence of the special model used by RiskMetrics, and it holds true because the model assumes that, on a very short time frame, the mean value of stock returns can be neglected, namely $\mu=0$. This, again, can be partially justified by looking at equation (1.15); if $T$ is small enough, we have that $\sqrt{T} \gg T$ and this implies that on a very short time frame, volatility dominates drift. The following example shows a simple computation of VaR using the RiskMetrics approach and the square root of time rule.
\begin{exmp}
The sample standard deviation of the continuously compounded
daily return of the German mark/U.S. dollar exchange rate was about 0.53\%
in June 1997. Suppose that an investor was long in \$10 million worth of
mark/dollar exchange rate contract. Then the 5\% VaR for a 1-day horizon of the
investor is
$$\$10,000,000 \times (1.65 \times 0.0053) = \$87,450$$
The corresponding VaR for 10-day horizon is
$$\$10,000,000 \times (\sqrt{10} \times 1.65 \times 0.0053) \approx \$276,541$$
Notice, however, that if we increase the time horizon the drift should not be neglected and the square root of time rule might produce poor estimations of VaR.
\end{exmp}
The RiskMetrics method for VaR calculation has some clear advantages, which made the approach so popular and widely used among financial operators. First of all, it is easy to understand and apply and produces reasonable estimations if correctly used. Another advantage is that it is an open method with a very well known computation procedure, it does not come in the form of a mysterious \vir{black box} like many other risk manager packages and this makes risk more transparent in the financial markets.

On the other hand, there are also several drawbacks. The first and perhaps most important one lies in the model assumptions: to assume that stock returns are normally distributed is generally unrealistic. The RiskMetrics group recommends the users to use a 95\% confidence level for computation and this produces reasonable estimations in general, but if we want to increase the confidence level, than we should be aware that real returns distributions have fat tails and this greatly influences the value of high-confidence quantiles. Indeed, we may notice that the 5\% quantile of a standard normal is approximately -1.65. It is very often found that despite the presence of fat tails, for many distributions the 5\% quantile is roughly -1.65. For example, the 95\% quantile of the Student t distribution with 7 degrees of freedom (which is fat tailed and has a kurtosis of 5 similar to the typical kurtosis of returns in financial markets) is -1.60, very close to -1.65. It is evident, however, that for higher significance levels (e.g. 99\%) the effect of fat tails becomes much stronger, and therefore the VaR will be seriously underestimated if one assumes normality. For example, the 1\% quantile of the Student t distribution considered above is -2.54, significantly larger than under the normality assumption (-2.33).

Moreover, as we have already anticipated, we should also be aware that if we consider large time horizon we should not neglect drift and the square root of time might underestimate the VaR.

\subsubsection{Parametric Approach}
The parametric approach consists in building a parametric statistical model describing portfolio returns, i.e. assuming a certain probability distribution describing them, and then fitting the model by using historical data. The parameters can be estimated using standard techniques, like maximum likelihood estimation. The ultimate goal of the parametric approach is to find \emph{analytic} expressions for VaR. Notice that the RiskMetric approach falls in this category as it assumes a multivariate normal distribution for returns. Closed-form expressions for VaR can be found also when returns have other distributions
 like the Student-t distribution.

\subsubsection{The Historical Method}
The historical method does not impose any distributional assumptions; the distribution of portfolio returns is constructed
from historical data. Hence, sometimes the historical simulation method is
called a nonparametric method. For example, the 99\% daily VaR of the
portfolio return is computed as the negative of the empirical 1\% quantile of
the observed daily portfolio returns. The observations are collected from a
predetermined time window such as the most recent business year.

While the historical method seems to be more general as it is free of any
distributional hypotheses, it has a number of major drawbacks. The most relevant one is that it implicitly
assumes  that  the  distribution  of  past  returns  is  a  good  and  complete  representation  of
expected future returns, which might be far from true. Secondly, it treats the observations as independent and identically distributed
(i.i.d.), which is not realistic. The daily returns data exhibits clustering
of the volatility phenomenon, autocorrelations and so on, which are
sometimes a significant deviation from the i.i.d. assumption. Moreover, recent returns should be considered more informative than old ones, while in the plain historical method all the observations are considered equally important. Finally, it is not reliable for estimation of VaR at very high confidence levels. A
sample of one year of daily data contains 250 observations, which is a
rather small sample for the purpose of the 99\% VaR estimation.
\subsubsection{Monte Carlo Simulation}
In contrast to the historical method, the Monte
Carlo method requires specification of a statistical model for the stocks
returns. The statistical model is multivariate, hypothesizing both the behavior of the stock returns on a stand-alone basis and their dependence. For
instance, the multivariate normal distribution assumes normal distributions for the stock returns viewed on a standalone basis and describes the
dependencies by means of the covariance matrix. The multivariate model
can also be constructed by specifying explicitly the one-dimensional distributions of the stock returns, and their dependence through a copula
function. 

The Monte Carlo method consists of the following basic steps:
\begin{enumerate}
\item \emph{Selection of a statistical model}. The statistical model should be
capable of explaining a number of observed phenomena in the data
such as heavy-tails, clustering of the volatility, and the like, which
we think influence the portfolio risk.
\item \emph{Estimation of the statistical model parameters}. A sample of
observed stocks returns is used from a predetermined time window,
for instance the most recent 250 daily returns.
\item \emph{Generation of scenarios from the fitted model}. Independent
scenarios are drawn from the fitted model. Each scenario is a
vector of stock returns that depend on each other according to the
presumed dependence structure of the statistical model.
\item \emph{Calculation of portfolio risk}. Compute portfolio risk on the basis
of the portfolio return scenarios obtained from the previous step
\end{enumerate}

The Monte Carlo method is a very general numerical approach to
risk estimation. It does not require any closed-form expressions and, by
choosing a flexible s tatistical model, accurate risk numbers can be obtained.

However, since Monte Carlo simulation requires users to make assumptions about the stochastic process describing portfolio returns, it  is  subject  to  model  risk.   It  also  creates  inherent  sampling  variability  because  of  the
randomization.  Different random numbers will lead to different results.  It may take a large number of iterations to converge to a stable VaR measure. Although, Monte  Carlo  Simulation  can  be  time-consuming  according  to  the  properties  of  problem, the main benefit of running it is that it can model instruments with non-linear and path-dependent payoff functions, especially complex derivatives.

\section{Expected shortfall}
Although it is rather easy to evaluate value at risk, in the previous section we have seen that VaR has a number of drawbacks as risk measure. The most important ones are that it is not sub-additive nor informative about the magnitude of the losses larger than VaR level.

To capture the behavior of the ``bad'' tail of the loss/payoff distribution, it is more sensible to use another risk measure related to the VaR, known as \emph{Expected Shortfall}. The expected short fall measures the weighted avarage loss that will occur if we lose more money than the $\alpha$-level VaR, where the weights are the probabilities associated to each loss larger than VaR. If we denote with $X$ the random payoff of our portfolio, this leads to the following mathematical formalization:
\begin{equation}
\mbox{ES}_{\alpha}(X) \doteq  -\frac{1}{\alpha} \int_0^{\alpha} \mbox{VaR}_p(X) \: \mathrm{d}p =  -\frac{1}{\alpha} \int_0^{\alpha} F_{X}^{-1}(\alpha)  \mathrm{d}p
\end{equation}
Notice that this can be equivalently written in terms of a conditional expectation:
\begin{equation}
\mbox{ES}_{\alpha}(X)= -\mathbb{E}[X | X < -\mbox{VaR}_{\alpha}(X)]
\end{equation}
the latter equality holds true if and only if the cumulative density function of $X$ is continuous at $x=\mbox{VaR}_{\alpha}(X)$. Notice that, by definition, it holds $\mbox{ES}_{\alpha}(X) \ge \mbox{VaR}_{\alpha}(X) \: \: \forall \: X, \: \forall \: \alpha \in [0,1]$.

It can be shown that ES satisfies all the axioms of coherent risk measures. One
consequence is that, unlike VaR, it is convex for all possible portfolios
which means that it always accounts for the diversification effect.

Closed-form expressions for ES can be found for some particular distributions, like the normal and the t-student distributions.
For real applications, it is useful to provide an estimation formula, that allows to evaluate ES from a sample of returns. Provided that we do not impose any distributional model, the ES of portfolio return can be estimated as follows. Denote the observed portfolio returns by $r_1,r_2,\dots,r_n$ at time instants $t_1,t_2,\dots, t_n$. Denote the sorted samples by $r_{(1)} \le r_{(2)} \le \dots \le r_{(n)}$. Thus $r_{(1)}$ equals the smallest portfolio return observed and $r_{(n)}$ is the largest one. The ES can be estimates according to the formula:
\begin{equation}
\widehat{ES}_{\alpha}(r)= -\frac{1}{\alpha} \left( \frac{1}{n} \sum_{k=1}^{\ceil{n\alpha}-1} r_{(k)}+\left(\alpha-\frac{\ceil{n\alpha}-1}{n} \right) r_{(\ceil{n\alpha})}\right)
\end{equation}
This formula can be usefully applied to estimate ES both from historical data and by means of Monte Carlo methods. If we want to use a monte carlo approach, we need a statistical model of returns. 
\subsection{Estimating Expected Shortfall}
The ideas behind the approaches of VaR estimation can be applied to ES.
We revisit the three methods from section 1.2 focusing on the
implications for ES.
\subsubsection{The multivariate normal assumption}
We have seen that, if we assume returns to be normally distributed like in the RiskMetrics approach, then the expected value and variance of the portfolio return $R_p$ can be expressed according to formulae (1.9) and (1.10).  Let us evaluate the expected shortfall of the portfolio under this assumption.
\begin{equation*}
\begin{split}
\mbox{ES}_{\alpha}(R_p) & = -\mathbb{E}[\left. R_p \right\vert R_p < -\mbox{VaR}_{\alpha}(R_p)] \\[2ex]
& = -\mathbb{E}\left[\left.R_p\right\vert  \frac{R_p-\mu_R}{\sigma_{R_p}} < -\frac{\mbox{VaR}_{\alpha}(R_p)+\mu_R}{\sigma_{R_p}} \right] 
\end{split}
\end{equation*}
Let us call $\frac{R_p-\mu_R}{\sigma_{R_p}} =Z \sim \mathcal{N}(0,1)$, so that $R_p=\sigma_{R_p} Z +\mu_R$. Moreover, by the property of VaR as risk measure we have that $\frac{\mbox{VaR}_{\alpha}(R_p)+\mu_R}{\sigma_{R_p}} = \mbox{VaR}_{\alpha}\left(\frac{R_p-\mu_R}{\sigma_{R_p}} \right)=\mbox{VaR}_{\alpha} (Z) =q_{1-\alpha}. $Then we have:
\begin{equation*}
\begin{split}
\mbox{ES}_{\alpha}(R_p) & = -\mathbb{E}\left[\sigma_{R_p} Z + \mu_R | Z < -q_{1-\alpha} \right]  \\[2ex]
& = -\sigma_{R_p} \mathbb{E}\left[Z | Z <- q_{1-\alpha} \right] - \mu_R \\[2ex]
& = -\frac{\sigma_{R_p}}{\alpha} \int_{-\infty}^{-q_{1-\alpha}} \frac{x}{\sqrt{2 \pi}} e^{ -\frac{x^2}{2}}  \: \mathrm{d}x -\mu_R \\[2ex]
&=  \frac{\sigma_{R_p}}{\alpha \sqrt{2 \pi}} \exp \left( -\frac{q_{1-\alpha}^2}{2} \right)-\mu_R \\[2ex]
&=C_{\alpha} \sigma_{R_p}-\mu_R
\end{split}
\end{equation*}
Here $C_{\alpha}$ is a constant whose value depends only on the confidence level $\alpha$ and can be calculated in advance. The above formula shows that ES can be calculated as the difference between the properly scaled standard
deviation of the random portfolio return and portfolio expected return. Notice that this is formally the same expression of the VaR under the same normality assumption. The only thing that differs is the value of the costant term $C_{\alpha}$, which in the case of VaR is simply the quantile $q_{\alpha}$ and we have $C_{\alpha} \ge q_{\alpha} \: \forall \: \alpha \in [0,1]$. Because of the simmetry of the normal distribution around the mean, the ES is symmetric as well under this assumption and does not make differences between long and short positions.

\subsubsection{Parametric Approach}
The parametric approach consists in building a parametric statistical model describing portfolio returns, i.e. assuming a certain probability distribution describing them, and then fitting the model by using historical data. The parameters can be estimated using standard techniques, like maximum likelihood estimation. The ultimate goal of the parametric approach is to find \emph{analytic} expressions for ES. Notice that the multivariate normal model falls in this category. Closed-form expressions for ES can be found also when returns have other distributions  like the Student-t distribution. 

\subsubsection{The historical method}
The historical method has several drawbacks mentioned in the previous section regarding VaR. We emphasize that it is very inaccurate for low-tail probabilities
such as 1\% or 5\%. Even with one year of daily returns, which amounts
to 250 observations, to estimate the ES at 1\% probability, we have to
use the three smallest observations which is quite insufficient. What makes
the estimation problem even worse is that these observations are in the
tail of the distribution; that is, they are the smallest ones in the sample.
The implication is that when the sample changes, the estimated ES may
change a lot because the smallest observations tend to fluctuate a lot.
\subsubsection{Monte Carlo simulation}
The basic steps of the Monte Carlo method are described in section 1.2.1. They are applied without modification. Essentially, we assume
and estimate a multivariate statistical model for the stocks return distribution. Thenwe sample from it, and we calculate scenariosfor portfolio return.
On the basis of these scenarios, we estimate portfolio ES using equation
(1.18) in which $r_1,r_2,\dots,r_n$ stands for the vector of generated scenarios.
Similar to the case of VaR, an artifact of the Monte Carlo method is
the variability of the risk estimate. Since the estimate of portfolio AVaR
is obtained from a generated sample of scenarios, by regenerating the
sample, we will obtain a slightly different value.

\chapter{Leverage and margining control}
Buying/short selling on margin has become an essential part of modern trading systems. In this chapter we  discuss how to manage this business from the point of view of the firm lending money to its customers.

\section{The simplest case: money reserve + one asset}
Say that a certain client wants to invest a fraction $w$ of his capital $C$ on a certain risky asset, whose daily return is given by the random variable $R$. The remaining fraction of the money, namely $1-w$, is kept by the client as a reserve. The firm allows the client to borrow money up to $l_R$ times the value of the original investment on the asset, so that 
with an upfront payment of $wC$ the client buys the stock for a value $l_RwC$.
The money borrowed by the client is then  $Cl_R w-Cw=Cw(l_R-1)$. 
The client portfolio at the end of the day has following (random) value:
\begin{equation}
\tilde C_+ = C \left( l_R w (1+R) +1-w\right).
\end{equation}
At the end of the day, however, the client must  pay back his debt, which means that the net amount remaining will be:
\begin{equation}
C_+ = \tilde C_+ -  Cw(l_R-1) = C\left( l_R w (1+R) +1-w-w(l_R-1) \right)= C ( l_R w R+1).
\end{equation}
It is convenient to express this quantity as a sort of return with respect to the money lended to the client. Thus, we define:
\begin{equation}
\Delta=\frac{l_R w R+1}{ l_R w}
\label{eq:Deltadef}
\end{equation}

Firm-wise, we want this $\Delta$ to be greater than a certain safety treshold $h$. and we are interested in finding a suitable range for the leverage factor $l_R$ such that this condition holds. There are at least two possible approaches.
\subsection{VaR-style approach}
A natural way of choosing the value of $l_R$ would be to ask that the probability for the client to go under the treshold $h$ is small enough. Namely, if we fix a confidence level $\alpha \in (0,1)$, we want that:
\begin{equation}
l_R \in \{ l_R>1 \: : \: \mathbb{P}(\Delta<h) \le \alpha \}				
\end{equation}
Plugging the expression of $\Delta$, we easily find that: 
\begin{equation}
\mathbb{P}(\Delta<0) = \mathbb{P}\left( R<h-\frac{1}{l_R w} \right)
\end{equation}
In order to carry out the calculation, we need a probabilistic model for the asset return $R$, which means that we have to specify a distribution for it. A common assumption, as we have seen, is to assume $R$ normally distributed, namely $R \sim \mathcal{N}(\mu_R, \sigma_R^2)$. In this case, by normalizing the variable $R$, we have:
\begin{align}
\mathbb{P}\left( \frac{R-\mu_R}{\sigma_R}<\frac{l_Rwh-1-l_R w \mu_R}{l_R w \sigma_R} \right) & \le \alpha \notag \\[2ex]
\mathbb{P}\left( Z<\frac{l_Rwh-1-l_R w \mu_R}{l_R w \sigma_R} \right) & \le \alpha  \notag \\[2ex]
\frac{l_Rwh-1-l_R w \mu_R}{l_R w \sigma_R} & \le q_{\alpha} \notag
\end{align}
and solving for $l_R$:
\begin{equation}
l_R \le \frac{1}{w (h-\mu_R-q_\alpha \sigma_R)}
\end{equation}
We have that $\mu_R+q_\alpha \sigma_R = \mu_R-q_{1-\alpha} \sigma_R=-\mbox{VaR}_{\alpha}(R)$, and plugging this in the above formula we finally obtain
\begin{equation}
l_R \le \frac{1}{w (h+ \mbox{VaR}_{\alpha}(R))}
\label{eq:leverage_varlimit}
\end{equation}
This formula bonds in a very simple way the maximum leverage factor with the asset riskiness, measured in this case by the VaR, and the amount of money invested on it.

We can observe that, since it must be $l_R\ge1$, we have
\begin{equation}
 \frac{1}{w (h+ \mbox{VaR}_{\alpha}(R))}\ge1 \quad \quad \Longrightarrow \quad \quad \mbox{VaR}_{\alpha}(R)\le -h-\frac{1}{w}
\end{equation}
This is a necessary condition in order to define the leverage factor. 
\begin{equation}
 1\le l_R \le \frac{1}{w (h+ \mbox{VaR}_{\alpha}(R))}	
\end{equation}

We can say that if the $l_R$ value resulting from (\ref{eq:leverage_varlimit})
is $< 1$, then it means that the asset is way too risky, and the firm shall not offer a leverage on it.

\medskip

Obvioulsy, a critical value for the treshold is $h=0$. If $\Delta$ falls beyond this value, it means that the client does not have enough money to pay back his debt. This framework can be used to set individual leverage factors for each asset and in this sense we usually set $w=1$ to be more conservative. 
\subsection{Expected shortfall approach}
The VaR-style approach that we have just discussed shares the same drawbacks with the concept of VaR, in particular it just tells us the probability that our client will not have the money to fully pay back his debt. A more interesting information for the firm would be to know the average magnitude of the client loss. 

This naturally brings to a ES-style approach; we are interested in keep the average loss per lended dollar
bounded by a certain threshold $h$.
\begin{equation}
 -\mathbb{E}[\Delta | \Delta \le 0] \le  h 
\end{equation}
plugging the expression of $\Delta$ we find
\begin{align}
 -\mathbb{E}\left[ l_R w R+1| R <-\frac{1}{l_Rw} \right] \le h l_R w \notag \\[2ex]
 -\mathbb{E}\left[ l_R w (\sigma_R Z+\mu_R)+1 | Z <-\frac{1+l_R w \mu_R}{l_R w \sigma_R} \right] \le  h l_R w \notag \\[2ex]
-l_R w \sigma_R \: \mathbb{E}\left[ Z| Z <-\frac{1+l_R w \mu_R}{l_R w \sigma_R} \right]-l_Rw\mu_R-1 \le  h l_R w \notag \\[2ex]
-\frac{l_R w \sigma_R}{\Phi \left(-\frac{1+l_R w \mu_R}{l_R w \sigma_R}\right)} \int_{-\infty}^{-\frac{1+l_R w \mu_R}{l_R w \sigma_R} } \frac{x}{\sqrt{2 \pi}} e^{ -\frac{x^2}{2}}  \: \mathrm{d}x -l_Rw\mu_R-1\le  h l_R w \notag \\[2ex]
\frac{l_R w \sigma_R}{\sqrt{2 \pi} \: \Phi\left(-\frac{1+l_R w \mu_R}{l_R w \sigma_R}\right)}  \exp \left( -\frac{1}{2} \left(-\frac{1+l_R w \mu_R}{l_R w \sigma_R}\right)^2 \right)-l_Rw\mu_R-1 \le  h l_R w \notag \\[2ex]
\frac{1}{\sqrt{2 \pi} \: \Phi\left(-\frac{1+l_R w \mu_R}{l_R w \sigma_R}\right)}  \exp \left( -\frac{1}{2} \left(-\frac{1+l_R w \mu_R}{l_R w \sigma_R}\right)^2 \right)-\frac{1+l_R w \mu_R}{l_R w \sigma_R} \le \frac{h}{\sigma_R} \notag
\end{align}
Let us put $x=\Phi\left(-\frac{1+l_R w \mu_R}{l_R w \sigma_R}\right)$, the above equation becomes
\begin{align}
\frac{\sigma_R}{\sqrt{2 \pi} x}  \exp \left( -\frac{1}{2} q_x^2 \right)+\sigma_R q_x \le  h \notag \\[2ex]
\frac{\sigma_R}{\sqrt{2 \pi} x}  \exp \left( -\frac{1}{2} q_{1-x}^2 \right)-\mu_R+\sigma_R q_x+\mu_R \le  h \notag \\[2ex]
\mbox{ES}_x(R)-\mbox{VaR}_{x}(R) \le h
\end{align}
This is a trascendental algebrical inequality that can be easily solved numerically. The corresponding inequality for $l_R$ is
\begin{equation}
l_R \le \frac{1}{-w (\mu_R+q_x^* \sigma_R)}
\end{equation}
where $q_x^*$ is the $x^*$- quantile of a standard normal and $x^*$ is the solution of the equation associated with inequality (3.10). This can be written as:
\begin{equation}
l_R \le \frac{1}{w\mbox{VaR}_{x^*}(R)}
\end{equation}
Inequality (3.10) can be seen also from the other way around: one could set a target shortfall probability $x$ and then obtain the associated everage loss. This equivalence between the two approaches comes from the normality assumption of returns. The two approaches are perfectly equivalent if we set $x=\alpha$.

\section{The general case}
In this part we consider a generic portfolio consisting of $n-1$ assets (labeled from 2 to $n$) and for each of them we assume to have already evaluated the associated leverage factor. Say that the client wants to add another asset to his portfolio (labeled as asset 1). The portfolio value at the end of the day will be:
\begin{equation}
\tilde C_+ =C \left(l_1w_1(1+R_1)+\sum_{k=2}^{n}l_kw_k(1+R_k)+1-\sum_{k=1}^n w_k \right)
\end{equation}
At the end of the day, the client will have to pay back his debt, which means that the net amount remaining, expliciting the unknown leverage factor $l_1$, will be:
\begin{equation}
C_+ = C \left( l_1w_1R_1+\sum_{k=2}^{n}l_kw_kR_k+1 \right)
\end{equation}
Moreover, let us call $\mathbf{l}_w$ the vector collecting the terms $l_k w_k$. Thus, the $\Delta$ can be written in vector form:
\begin{equation}
\Delta=\frac{\mathbf{l}_w^T \mathbf{R}+1}{\mathbf{l}_w^T \mathbbm{1}}
\end{equation}
\subsection{VaR-style approach}
Fixed the confidence level $\alpha \in [0,1]$, we are interested in evaluating for which values of $l_1$ it holds
\begin{equation}
\mathbb{P}(\Delta<0) \le \alpha
\end{equation}
Let us assume that stock returns have a multivariate normal distribution. In that case we have that $\mathbb{E}[\mathbf{l}_w^T \mathbf{R}]=\mathbf{l}_w^T \boldsymbol\mu$ and $\mathrm{Var}(\mathbf{l}_w^T \mathbf{R})= \mathbf{l}_w^T \Sigma \mathbf{l}_w$. Plugging the expression of $\Delta$ and normalizing the variable, we have that
\begin{align}
\mathbb{P}(\mathbf{l}_w^T \mathbf{R}<-1) & \le \alpha \\[2ex]
\mathbb{P}\left(Z<-\frac{1+\mathbf{l}_w^T \boldsymbol\mu}{ \sqrt{\mathbf{l}_w^T \Sigma \mathbf{l}_w}} \right) & \le \alpha \notag \\[2ex]
-\frac{1+\mathbf{l}_w^T \boldsymbol\mu}{ \sqrt{\mathbf{l}_w^T \Sigma \mathbf{l}_w}} & \le q_{\alpha} \notag \\[2ex]
q_{\alpha}\sqrt{\mathbf{l}_w^T \Sigma \mathbf{l}_w} + \mathbf{l}_w^T \boldsymbol\mu & \ge -1
\end{align}
and rememebering that $q_{\alpha}=-q_{1-\alpha}$, the latter formula can be written as
\begin{equation}
\mbox{VaR}_{\alpha}(\mathbf{l}_w^T \mathbf{R}) \le 1
\end{equation}
The latter inequality must be solved for variable $l_1$, since we assume that all the other factors have already been determined. The value for which the equality holds $l_1^*$ will be the maximum leverage factor associated to the corresponding asset that the firm is willing to offer. Thus, the client can borrow money to be invested on this asset up to $l_1^* w$. If the client decides to invest less money, say $l_1' w$, with $l_1'<l_1^*$,  then $l_1'$ will be the leverage factor to be used in equation (3.20) when other leverage factors have to be determined for other assets.

\subsubsection{Leverage optimization}
The VaR-style approach can be generalized in the case we have to set more than one leverage factor at the same time (for example, when the client buys more assets at the same time). This is useful in particular for the client who can rebalance his factors in a clever way by keeping his default risk under control. In particular, our goal is to obtain the maximum leverage factors without violating the VaR constraint (3.20). A possible formulation of this problem is the following
\[
\begin{cases}
\underset{\mathbf{l}_w}\max \quad \min\mathbf{l}_w \\[2ex]
\: \: \text{s.t.} \quad  \mbox{VaR}_{\alpha}(\mathbf{l}_w^T \mathbf{R}) \le 1
\end{cases}
\]
We could also maximize the avarage factor:
\[
\begin{cases}
\underset{\mathbf{l}_w}\max \quad  \norm{\mathbf{l}_w} \\[2ex]
\: \: \text{s.t.} \quad  \mbox{VaR}_{\alpha}(\mathbf{l}_w^T \mathbf{R}) \le 1
\end{cases}
\]
We may also want to maximize the avarage value of the client porfolio at the end of the day, which brings to the following problem:
\begin{equation}
\begin{cases}
\underset{\mathbf{l}_w}\max \quad  \mathbf{l}_w^T \boldsymbol\mu \\[2ex]
\: \: \text{s.t.} \quad  \mbox{VaR}_{\alpha}(\mathbf{l}_w^T \mathbf{R}) \le 1
\end{cases}
\end{equation}
Since we are dealing with normally distributed returns, we know that VaR is a convex function, which in turn implies that the above problems can be cast as a convex optimization problems and so they admit a unique global solution.
\subsection{Expected shortfall approach}
In this case we are interested in keeping the potential loss bounded by a threshold $h$.
\begin{equation}
 -\mathbb{E}[\Delta | \Delta <0] \le  h
\end{equation}
plugging the expression of $\Delta$ we find
\begin{equation}
 -\mathbb{E}[\mathbf{l}_w^T \mathbf{R}+1 | \mathbf{l}_w^T \mathbf{R}<-1] \le \mathbf{l}_w^T \mathbbm{1} h
\end{equation}
From this point the passages are the same of the previous section. At the end, we have the following inequality:
\begin{equation}
\frac{\sqrt{\mathbf{l}_w^T \Sigma \mathbf{l}_w}}{\sqrt{2 \pi} \: \Phi\left(-\frac{1+\mathbf{l}_w^T \boldsymbol\mu}{ \sqrt{\mathbf{l}_w^T \Sigma \mathbf{l}_w}}\right)}  \exp \left( -\frac{1}{2} \left(-\frac{1+\mathbf{l}_w^T \boldsymbol\mu}{ \sqrt{\mathbf{l}_w^T \Sigma \mathbf{l}_w}}\right)^2 \right)-1-\mathbf{l}_w^T \boldsymbol\mu \le\mathbf{l}_w^T \mathbbm{1}  h
\end{equation}
and putting $x=\Phi \left( -\frac{1+\mathbf{l}^T \boldsymbol\mu}{ \sqrt{\mathbf{l}^T \Sigma \mathbf{l}}} \right)$ the inequality becomes:
\begin{align}
\frac{\sqrt{\mathbf{l}_w^T \Sigma \mathbf{l}_w}}{\sqrt{2 \pi} x}  \exp \left( -\frac{1}{2} q_x^2 \right)+\sqrt{\mathbf{l}_w^T \Sigma \mathbf{l}_w} \: q_x \le \mathbf{l}_w^T \mathbbm{1} h \notag \\[2ex]
\frac{\sqrt{\mathbf{l}_w^T \Sigma \mathbf{l}_w} }{\sqrt{2 \pi} x}  \exp \left( -\frac{1}{2} q_{1-x}^2 \right)-\mathbf{l}_w^T \boldsymbol\mu+\sqrt{\mathbf{l}_w^T \Sigma \mathbf{l}} \: q_x+\mathbf{l}_w^T \boldsymbol\mu \le\mathbf{l}_w^T \mathbbm{1}  h \notag \\[2ex]
\mbox{ES}_x(\mathbf{l}_w^T \mathbf{R})-\mbox{VaR}_{x}(\mathbf{l}_w^T \mathbf{R}) \le\mathbf{l}_w^T \mathbbm{1} h
\end{align}
and the latter inequality is formally equivalent to (3.10), which was derived for the simple case with one asset only. Likewise, once we have found the solution $x^*$ of the equation associated to the above inequality, the inequality for $\mathbf{l}$ reads as:
\begin{align}
q_{x^*}\sqrt{\mathbf{l}_w^T \Sigma \mathbf{l}_w} + \mathbf{l}_w^T \boldsymbol\mu & \ge -1 \notag \\
\mbox{VaR}_{x^*}(\mathbf{l}_w^T \mathbf{R}) & \le 1
\end{align}
which must be solved for the unknown $l_1$. Again, if we set $x^*=\alpha$, the VaR and ES approaches are perfectly equivalent.

Notice that, likewise the VaR approach, if we want to define more than one leverage factors at the same time we could solve the following optimization problem: 
\[
\begin{cases}
\underset{\mathbf{l}_w}\max \quad \mathbf{l}_w^T \boldsymbol\mu \\[2ex]
\: \: \text{s.t.} \quad  \frac{\sqrt{\mathbf{l}_w^T \Sigma \mathbf{l}_w}}{\sqrt{2 \pi} \: \Phi\left(-\frac{1+\mathbf{l}_w^T \boldsymbol\mu}{ \sqrt{\mathbf{l}_w^T \Sigma \mathbf{l}_w}}\right)}  \exp \left( -\frac{1}{2} \left(-\frac{1+\mathbf{l}_w^T \boldsymbol\mu}{ \sqrt{\mathbf{l}_w^T \Sigma \mathbf{l}_w}}\right)^2 \right)-1-\mathbf{l}_w^T \boldsymbol\mu \le\mathbf{l}_w^T \mathbbm{1}  h
\end{cases}
\]
Since the expected shortfall is a convex function, this is again a convex optimization problem. 

\section{Simulations}

In this section we use the results derived in the previous parts, and in particular the VaR-approach, to analyze a simple portfolio consisting of five assets. More in details, we have considered a four years-time frame for each asset's return; the first two years of this time frame have been used to calibrate the model and the last ones for the actual backtesting. In the following we have set $\alpha=5 \cdot 10^{-7}$ and $h=0$. Table 3.1 shows assets and the associated weights considered for the protfolio.
\begin{table}[H]
\centering
\label{my-label}
\begin{tabular}{|c|c|c|c|c|}
\hline
\textbf{Asset} & \textbf{\begin{tabular}[c]{@{}c@{}}Portfolio \\ Weights\end{tabular}} & \textbf{\begin{tabular}[c]{@{}c@{}}Maximum\\Sequential \\ Leverage \\ Factor\end{tabular}} & \textbf{\begin{tabular}[c]{@{}c@{}}Leverage\\ Factor used\\ by client\end{tabular}} & \textbf{\begin{tabular}[c]{@{}c@{}}Optimized\\ Leverage\\ Factor\end{tabular}} \\ \hline
DiaSorin & 0.1 & 136.5449 & 100 & 44.8860 \\ \hline
Tiscali & 0.2 & 18.8557 & 16 & 13.8921 \\ \hline
Generali & 0.2 & 7.8106 & 3 & 14.0963 \\ \hline
Geox & 0.3 & 3.4976 & 1 & 6.1657 \\ \hline
FCA & 0.1 & 6.7134 & 6.7134 & 30.7780 \\ \hline
\end{tabular}
\caption{Portfolio weights and leverages.}
\end{table}
Let us  simulate a possible trading strategy pursued by a client. Say that such a client wants to invest a fraction $w_1=0.1$ of his wealth on DiaSorin. By using formula (3.7) the firm evaluates the maximum associated leverage factor: $l_1^*=136.5449$. Let us imagine an extreme situation for which the client decides to use $l_1=100$. Now he still has some \vir{potential leverage} to use on other assets. For instance, suppose that the client invests $w=0.2$ on Tiscali. Again, the firm can evaluate the next sequential maximum leverage factor, but this time, since we are dealing with a portfolio consisting of two possibly correlated assets, we use formula (3.19) to evaluate it. Notice that the unknown in such formula is $l_2$ and we put $l_1=100$. Carrying out the calculations we obtain $l_2^*= 18.8557$. This is the maximum leverage factor that the client can still use; the client decides to use $l_2=16$. We can do the same thing for the remaining assets; resuts are showed in table 3.1.

Notice that the last factor associated to FCA is equal to the maximum one and thus it saturates the constraint (3.20). Infact, by running a monte carlo simulation assuming that returns follow a multivariate normal distribution, we can see that the $\alpha$-quantile of the portfolio return distribution measured is $q_{\alpha} \approx -0.993$, which is not surprisingly very close to the theorical value -1. 
\begin{figure}[H]
\center
\includegraphics[scale=0.8]{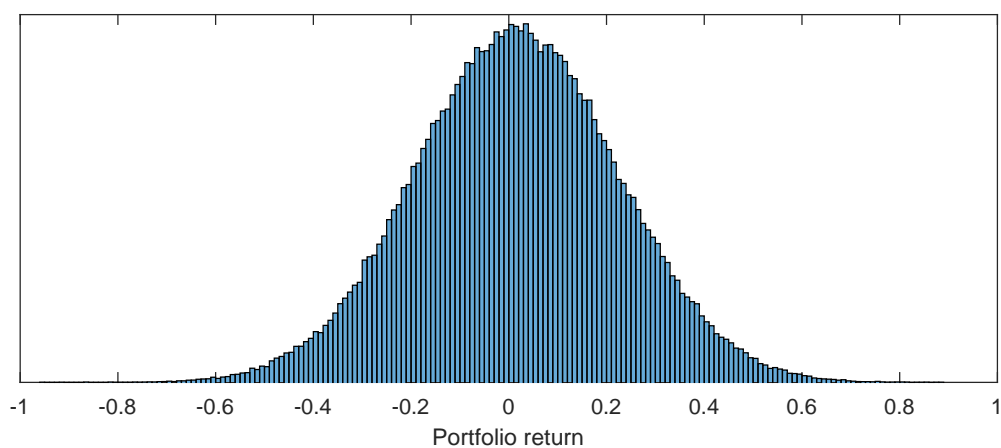}
\caption{Portfolio return (normally distributed returns, 300000 samples).}
\label{fig03}
\end{figure}
It is much more interesting, however, to backtest the results of the model using real historical returns. The empirical portfolio return distribution is showed in figure 3.2.
\begin{figure}[H]
\center
\includegraphics[scale=0.8]{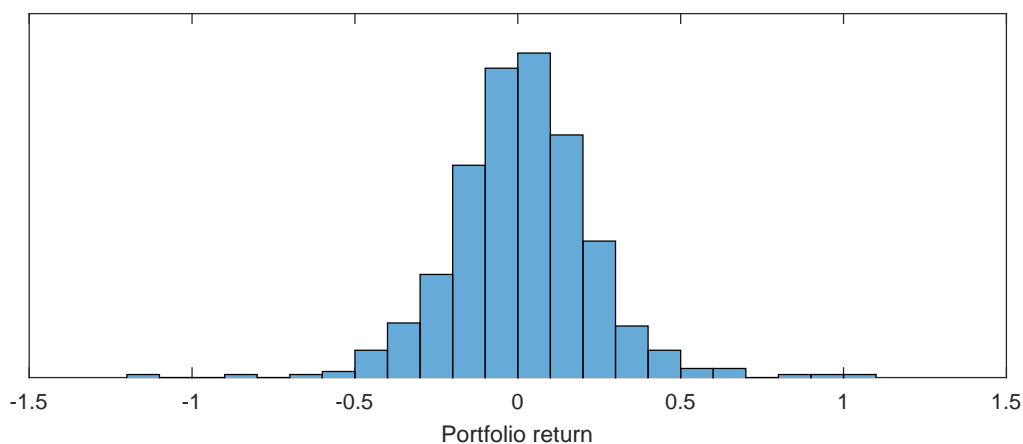}
\caption{Portfolio return (historical returns, 505 observations).}
\label{fig03}
\end{figure}
In this case, the empirical quantile observed  turns out to be $q_{\alpha}=-1.274$; since this value is larger than -1, the constraint of the problem (3.20) is actually violated. This is not surprising in this case, indeed in this example we solved equation (3.20) using the normality assumption of returns while it is well known that real returns, used for backtesting, do not follow a normal distribution and exhibit fat tails. More refined techniques shall be used to evaluate the VaR in order to have conservative results.

\medskip

The client could also use the \vir{clever} strategy that we discussed in section (3.2.1). Knowing in advance the assets he wants to invest on, he may plan what leverage factors to use by solving the problem (3.21). The optimized leverage factors are shown in table 3.1.

%    This factors are always feasible, in the sense that they are always smaller than the maximum leverage factors that we evaluate with formula (3.19). For instance, we see that the optimal leverage factor of Pininfarina is $l_1^o=20.6733$. Using this value in formula (3.19), we find that the maximum leverage factor associated, for example, to Tiscali is $l_2^*=28.8336 > l_2^o=11.8321$. If we use the two optimal factors for Pininfarina and Tiscali, we find that the maximum factor available for investing on Generali is $l_3^*=55.6641>l_3^o =12.0372$ and so on.

Figure 3.3 shows the empirical distribution of the client's portfolio return if he used such optimized factors.
\begin{figure}[H]
\center
\includegraphics[scale=0.8]{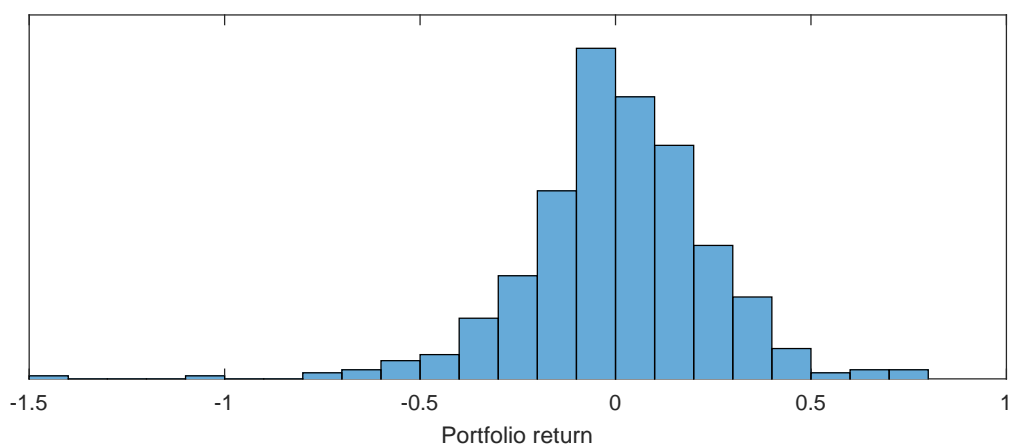}
\caption{Portfolio return with optimized leverage  (historical returns, 505 observations).}
\label{fig03}
\end{figure}
The empirical $\alpha$-quantile of the above distribution turns out to be $q_{\alpha}=-1.374$; since this value is larger than -1, the constraint of the problem (3.21) is actually violated. Again this is not surprising for the same reasons stated before.

  For both strategies, we can observe how the leverage magnifies both positive and negative returns. Nevertheless, simulations show that the model, even under the simplified hypothesis of normal returns, manages to keep the risk under control from the point of view of the firm by minimizing the probability for the client to default.

\medskip

 By running some tests, we can also derive the leverage factors required by the package ETMS. From the leverage factors we can obtain the corresponding confidence levels of the VaR approach. Results are shown in the following table:

\begin{table}[h]
\centering
\label{my-label}
\begin{tabular}{|c|c|c|c|c|c|}
\hline
\textbf{Asset} & \textbf{\begin{tabular}[c]{@{}c@{}}Directa\\  leverage factors\end{tabular}} & \textbf{\begin{tabular}[c]{@{}c@{}}ETMS \\ leverage factor\end{tabular}} & \textbf{\begin{tabular}[c]{@{}c@{}}Equivalent $\alpha$-level\\  (ETMS)\end{tabular}} & \textbf{\begin{tabular}[c]{@{}c@{}}Avarage\\  daily return\end{tabular}} & \textbf{Daily volatility} \\ \hline
FCA & 5 & 7.5998 & 9.50E-09 & 1.29E-03 & 0.0237 \\ \hline
Generali & 5 & 10.9424 & 2.04E-09 & 4.35E-04 & 0.0156 \\ \hline
Pininfarina & 2 & 3.844 & 9.72E-16 & 2.19E-03 & 0.0331 \\ \hline
Tiscali & 2 & 3.0098 & 1.93E-25 & 1.68E-03 & 0.0322 \\ \hline
ENI & 5 & 9.9933 & 1.34E-07 & -7.43E-05 & 0.0194 \\ \hline
\end{tabular}
\caption{ETMS package}
\end{table}

\newpage

\section{Margining control - Single asset case}
Say that a client, with an initial capital $C$, wants to invest a certain amount of money $w>C$  on a risky asset, whose daily return is given by the random variable $R$ and the associated margining factor is $a$ (where obviously $0<a<1$). The investment is possible since it is done on margin. Indeed, the liquidity of the client after the investment is $L=C-w<0$; notice that this quantity is exactly the amount of money borrowed by the client. The so called \emph{Marginal Availability} is the \vir{virtual} amount of money that is still available for the client after depositing the margin for his investment: $M_0=C-aw$. Notice that this quantity must be positive in order to carry out the trade and this implies that the maximum amount of money that can be invested at first on that asset is $w=\frac{c}{a}$.

\smallskip

What is the situation at the end of the day? The margin to be deposited is updated considering the closing price of the asset, which means that the marginal availability at the EOD becomes:
\begin{equation}
M=C-w+(1-a)(1+R)w
\end{equation}
Firm-wise, we want this random variable to be positive, otherwise the client (or the firm) must close the position or deposit more money in order to restore the margin. If the loss on $M$ is too large, the client may not be able to pay back is debt.

In order to be conservative, we consider the most risky situation, for which the client invests the maximum amount available: $w=\frac{c}{a}$. With this choice, the formula (3.27) becomes:
\begin{equation}
M_m=RC\left(\frac{1}{a}-1\right)
\end{equation}
In order to pin down a suitable value for $a$, we ask that the probability for the marginal availability to fall beyond a certain treshold, expressed as a fraction fo the initial capital $C$, is less than $\alpha$. In formule, we want that:
\begin{equation}
\mathbb{P}\left( C\frac{1-a}{a}R_P \le-hC \right) \le \alpha \quad \quad \Longrightarrow \quad \quad \mathbb{P}\left(R_P \le-h \frac{a}{1-a} \right) \le \alpha
\end{equation}
The above formula implies the following:
\begin{equation}
-h \frac{a}{1-a} \le -\mbox{VaR}_{\alpha} (R)
\end{equation}
and solving with respect to $a$ we find:
\begin{equation}
a \ge \frac{\mbox{VaR}_{\alpha} (R)}{h+\mbox{VaR}_{\alpha} (R)}
\end{equation}
This formula provides a simple connection between the value at risk of the asset and the associated margining factor.
As we could have expected, we have that $a \to 1$ when $h \to 0$ or $\mbox{VaR}_{\alpha} (R) \to \infty$.

In figure 3.4 we plotted the value of the margin factor as a function of the parameter $h$ for three different assets.
\begin{figure}[ht]
\center
\includegraphics[scale=0.8]{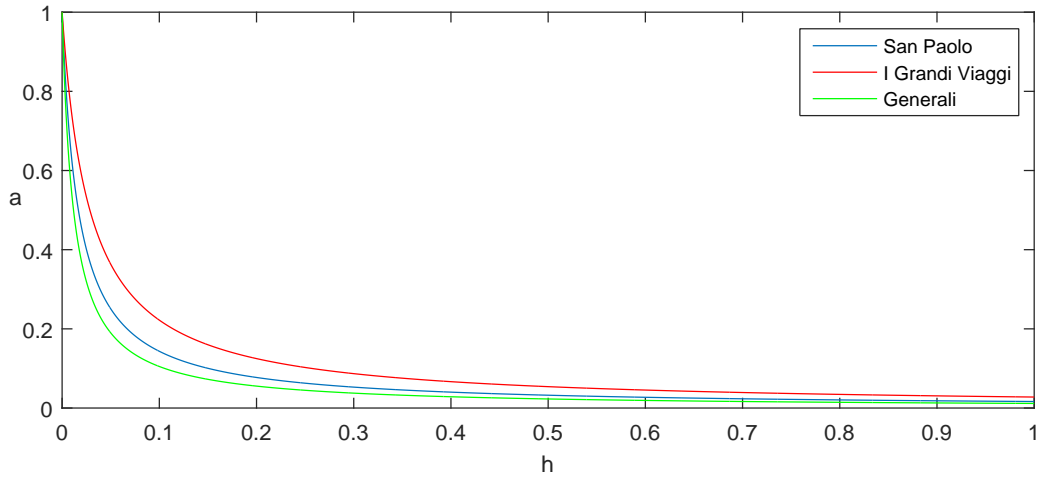}
\caption{Margining factor $a$ for some assets}
\label{fig03}
\end{figure}
\section{The general case}
The argument that we have used for the single asset margining factor can be easly generalized for a complex portfolio.

In this general case, using vector notation, the initial marginal availability can be written as: $M_0=C-a \mathbf{w} \mathbbm{1}$; notice that here $a$ is the overall margining factor of the portfolio. At the EOD, the marginal availability will be;
\begin{equation}
M=C-\mathbf{w} \mathbbm{1}+(1-a)(\mathbbm{1}+\mathbf{R})^T \mathbf{w} 
\end{equation}
This can be equivalently re-written as:
\begin{equation}
M=(1-a) \mathbf{w}^T\mathbf{R}  +C-a \mathbf{w}^T \mathbbm{1}
\end{equation}
Like before, we want to evaluate a suitable value for $a$ considering the most risky situation, that is, when the client invests all the available marginal availability. Notice that, in this case, this means;
\begin{equation}
 \mathbf{w}^T \mathbbm{1}=\frac{C}{a}
\end{equation}
so that the sum of the invested money saturates $M_0$. Under this circumstance, we can always write the following:
\begin{equation}
 \mathbf{w} =\frac{C}{a} \mathbf{x}  \quad \quad  \mbox{where} \quad \quad  \mathbf{x}^T \mathbbm{1} =1
\end{equation}
The vector $\mathbf{x}$ represents the weights of each asset in the client portfolio and it is a known quantity. Plugging the above formula in (3.34) we obtain:
\begin{equation}
M_m=C\frac{1-a}{a} \mathbf{x}^T\mathbf{R} 
\end{equation}
Notice that $\mathbf{x}^T\mathbf{R} =R_P$ is just the daily return of the client portfolio.

In order to pin down a suitable value for $a$, we ask that the probability for the marginal availability to fall beyond a certain treshold, expressed as a fraction fo the initial capital $C$, is less than $\alpha$. In formule, we want that:
\begin{equation}
\mathbb{P}\left( C\frac{1-a}{a}R_P \le-hC \right) \le \alpha \quad \quad \Longrightarrow \quad \quad \mathbb{P}\left(R_P \le-h \frac{a}{1-a} \right) \le \alpha
\end{equation}
The above formula implies the following:
\begin{equation}
-h \frac{a}{1-a} \le -\mbox{VaR}_{\alpha} (R_P)
\end{equation}
and solving with respect to $a$ we find:
\begin{equation}
a \ge \frac{\mbox{VaR}_{\alpha} (R_P)}{h+\mbox{VaR}_{\alpha} (R_P)}
\end{equation}
This formula provides a simple connection between the value at risk of the portfolio and the overall margining factor to be used.

Notice that, according to formula (3.39), a well diversified portfolio reduces the VaR and so the margining factor (notice that this is true given that the portfolio return follows an elliptical ditribution). The following example depicts a possible use of this model.
\begin{exmp}
Say that a client has an initial capital of $C= \$ 10000$ and he decides to invest a total amount of $W=\$30000$ on the following assets: $w_1=\$6000$ on \emph{Intesa San Paolo}; $w_2=\$21000$ on \emph{I Grandi Viaggi}; $w_3=\$3000$ on \emph{Generali}.

From this, we can easly obtain the weights of the client portfolio:
 \begin{equation}
     \mathbf{x}=\frac{1}{W}\begin{bmatrix}
         w_1 \\
         w_2 \\
         w_3
        \end{bmatrix}=\begin{bmatrix}
         0.2 \\
         0.7 \\
         0.1
        \end{bmatrix}
  \end{equation}
The next step consists in evaluating the VaR of the portfolio ($\alpha=0.001$). We assume that asset returns follow a normal distribution, fitted by using historical data of the last two years (this assumption is not necessary, we can evaluate the portfolio VaR in more refined ways). We can then estimate the VaR:
\begin{equation}
\mbox{VaR}_{\alpha}(R_P) =\mbox{VaR}_{\alpha}(0.2 R_1+0.7R_2+0.1R_3) = 0.0804
\end{equation}
Finally, choosing for example $h=0.2$, we can evaluate the minimum margining factor by applying formula (3.39):
\begin{equation}
a^*=  \frac{0.0804}{0.2+0.0804}=0.2867
\end{equation}
With this value we can obtain the marginal availability:
\begin{equation}
M_0=C-a^*W= 10000-0.2867 \times 30000=  \$ 1399.336
\end{equation}
Since this vaue is positive, the order can be executed. 

\medskip

Say now that the client wants to add another asset in his portfolio, for example he wishes to invest another $w_4=\$10000$ on \emph{ENI}. Notice that this amount is way larger than the current marginal availability. Shall we allow this trade? To answer this question, we need to check the riskiness of the new portfolio and evaluate the associated value for the new margining factor. Indeed, the new position may reduce or increase the overall riskiness.

  The total amount invested becomes $W=40000$ and thus the new portfolio weights are: 
 \begin{equation}
     \mathbf{x}=\frac{1}{W}\begin{bmatrix}
         w_1 \\
         w_2 \\
         w_3 \\
         w_4
        \end{bmatrix}=\begin{bmatrix}
         0.150 \\
         0.525 \\
         0.075 \\
         0.250
        \end{bmatrix}
  \end{equation}
From this we find the new portfolio VaR:
\begin{equation}
\mbox{VaR}_{\alpha}(R_P) =\mbox{VaR}_{\alpha}(0.15 R_1+0.525 R_2+0.075 R_3+0.25 R_4) =0.0663
\end{equation}
Notice that the VaR has been reduced by the new position (this is not surprising as the portfolio is more diversified). The new margining factor is:
\begin{equation}
a^*=  \frac{0.0663}{0.2+0.0663}=0.2491
\end{equation}
Finally, we can re-evaluate the marginal availability considering the new amount invested and the new margining factor:
\begin{equation}
M_0=C-a^*W=  10000-0.2491 \times 40000= \$ 36.3808
\end{equation}
Since this value is positive, we can allow the trade.

Notice, for example, that if the new amount invested on ENI was $w_4=\$15000$, we would obtain an even lower VaR and margining factor $a^*=0.2378$ but a negative marginal availability: $M_0=-\$700.9631$ and so we should deny the trade. 
\end{exmp}
The above example shows how this margining system takes into account the correlation between assets and the diversification effect. Indeed, one may also compute a single margining factor for each new asset added to the portfolio. Indeed, if we consider the generic $k$-th asset of a portfolio, we may evaluate:
\begin{equation}
a_k^*= \frac{\mbox{VaR}_{\alpha} (R_k)}{h+\mbox{VaR}_{\alpha} (R_k)}
\end{equation}
It is easy to show that the overall margining factor for a portfolio consisting in $n$ assets is a weighted avarage of the single margining factors, where the weights are the sums of money invested on each asset:
\begin{equation}
a^*= \frac{\sum_{k=1}^n w_k a^*_k}{\sum_{k=1}^nw_k}= \frac{\sum_{k=1}^n w_k a^*_k}{W}=\sum_{k=1}^n x_k a^*_k
\end{equation}

\medskip
Formula (3.39) can be used also with a portfolio containing derivatives. In this case however, the evaluation of the portfolio VaR needs some extra care as the relationship between prices of the assets and the corresponding value of the portfolio becomes non linear (this is due to the non linear payoff of an option). There are many ways to evaluate the VaR for such portfolios analitically, for example the delta-gamma approximation. A more accurate way, but also computationally demanding, is to run a monte carlo simulation of the portfolio return. 

Adding options to a portfolio can be a useful strategy to hedge part of the risk away and, consequently, to lower the margin factor. However, it can also increase the portfolio riskiness considerably. The following example shows how the margin factor changes when adding a simple vanilla option to the portfolio.
\begin{exmp}
Like in the previous example, say that a client has an initial capital of $C= \$ 10000$ and he decides to invest a total amount of $W=\$30000$ on the following assets: $w_1=\$6000$ on \emph{Intesa San Paolo}; $w_2=\$21000$ on \emph{I Grandi Viaggi}; $w_3=\$3000$ on \emph{Generali}.

We have already seen that the margining factor for such portfolio is:
\begin{equation}
a^*=0.2867
\end{equation}
With marginal availability:
\begin{equation}
M_0=C-a^*W= 10000-0.2867 \times 30000=  \$ 1399.336
\end{equation}

\medskip

Say now that the client wants to invest $w_4=\$10000$ on vanilla european-style \emph{put} options written on I Grandi Viaggi expiring in 10 months and with strike price $K$ equal to the last observed price of the underlying: $K=\$0.85$  \footnote{The Black-Scholes formula has been used to price the option. We have used the following data: annualized risk-free rate $r=0.10$; annualized volatility of the asset $\sigma=\sigma_d \sqrt{252}$ where $\sigma_d$ is the daily volatility and we consider 252 trading days per year.}. Shall we allow this trade? To answer this question, we need to check the riskiness of the new portfolio and evaluate the associated value for the new margin factor. 
  The total amount invested becomes $W=40000$ and thus the new portfolio weights are: 
 \begin{equation}
     \mathbf{x}=\frac{1}{W}\begin{bmatrix}
         w_1 \\
         w_2 \\
         w_3 \\
         w_4
        \end{bmatrix}=\begin{bmatrix}
         0.150 \\
         0.525 \\
         0.075 \\
         0.250
        \end{bmatrix}
  \end{equation}
By means of monte carlo simulation, we estimate the new VaR of the portfolio
\begin{equation}
\mbox{VaR}_{\alpha}(R_P) =\mbox{VaR}_{\alpha}(0.15 R_1+0.525 R_2+0.075 R_3+0.25 R_{put}) =0.0123
\end{equation}
Where $R_{put}$ is the return of the put option.
Notice that the VaR has been significantly reduced by the new position. This is not surprising, indeed taking a long position on a put while having a long position on the underlying is a hedge strategy known as \emph{protective put}. The new margining factor is:
\begin{equation}
a^*=  \frac{0.0123}{0.2+0.0123}=0.0609
\end{equation}
Finally, we can re-evaluate the marginal availability considering the new amount invested and the new margining factor:
\begin{equation}
M_0=C-a^*W=  10000-0.0609 \times 40000= \$7563.5
\end{equation}
Since this value is positive, we can allow the trade. Notice that in this case the marginal availability has been increased by adding the new position.

\medskip

The situation would have been completely different if we added a \emph{call} option to the portfolio with same strike price and maturity as before. Say that the client invests $w_4=\$2000$ on call options written again on IGV. In this case we find:
\begin{equation}
\mbox{VaR}_{\alpha}(R_P) =0.0950
\end{equation}
\end{exmp}
and
\begin{equation}
a^*=  \frac{0.0950}{0.2+0.0950}=0.3218
\end{equation}
Notice that the VaR, and consequently the margining factor have increased. The corresponding marginal availability is:
\begin{equation}
M_0=C-a^*W=  10000-0.3218 \times 32000= -\$307.71
\end{equation}
Since this value is negative we shall not allow the trade.

\begin{exmp}
In this example we shall simulate a two-day trading scenario with margining control. We have considered a four years-time window for each asset return. The first two years have been used to evaluate the VaR while the other two for backtesting.

\smallskip

Let us consider a client having an initial capital of $C= \$ 10000$. She decides to invest a total amount of $W=\$25000$ on the following assets: $w_1=\$7500$ on \emph{FCA}; $w_2=\$7500$ on \emph{Monte Dei Paschi}; $w_3=\$10000$ on \emph{Diasorin}.

Carrying out the calculations (using $h=0.1$ and $\alpha=0.001$), we see that the margin factor for such portfolio is:
\begin{equation}
a_1^*=0.3977
\end{equation}
With marginal availability:
\begin{equation}
M_1^i=C-a_1^*W= 10000-0.3977 \times 25000=  \$ 56.901
\end{equation}
Now we use the historical returns of the last two years to simulate the situation at the end of the day. Figure 3.2 shows the empirical distribution of the marginal availability at EOD. 
Notice that, since the marginal availability is almost saturated by the amount invested by the client, we should expect the empirical $\alpha$-quantile of the marginal availability distribution divided by C to be close to $h=0.1$. Indeed, the empirical value is $h_{emp}=-q_{\alpha}(M)/C=0.1167$.

Figure 3.3 shows the empirical distribution of the portfolio final value at the end of day one.
\begin{figure}[H]
\center
\includegraphics[scale=0.8]{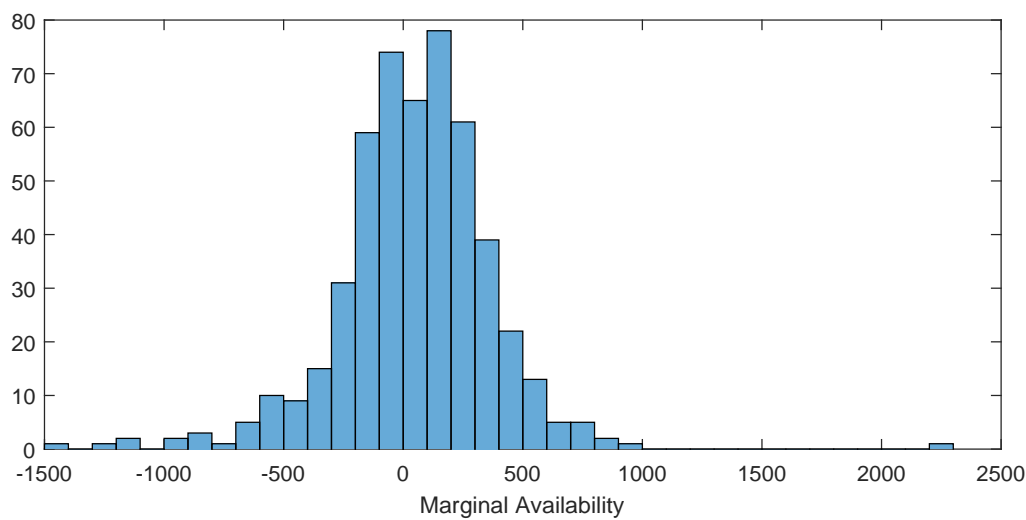}
\caption{Marginal availability distribution at the end of day one.}
\label{fig03}
\end{figure}

\begin{figure}[H]
\center
\includegraphics[scale=0.8]{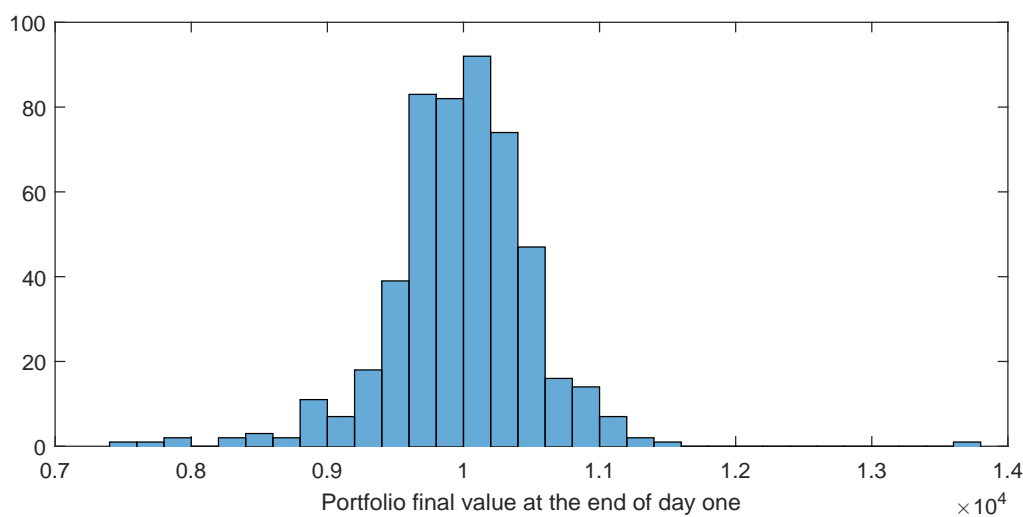}
\caption{Portfolio final value distribution at the end of day one.}
\label{fig03}
\end{figure}

\medskip
Let us now consider a possible scenario at the end of day one. We pick a realisation of asset returns among the historical ones used for backtesting. Say that with such returns the actual marginal availability at the end of day one is $M_1^f=\$ 255.12$ and the new portfolio value is $C_1^f=\mathbf{w}^T\mathbf{R}+C=\$ 10329$.

Say now that the client wants to invest $w_4=\$8500$ on vanilla european-style \emph{put} options written on Monte dei Paschi expiring in 10 months and with strike price $K$ equal to the last observed price of the underlying: $K=\$0.2795$  \footnote{The Black-Scholes formula has been used to price the option. We have used the following data: annualized risk-free rate $r=0.10$; annualized volatility of the asset $\sigma=\sigma_d \sqrt{252}$ where $\sigma_d$ is the daily volatility and we consider 252 trading days per year.}. Shall we allow this trade? To answer this question, we need to check the riskiness of the new portfolio and evaluate the associated value for the new margin factor. 
  The total amount invested becomes $W_2= \$ 33500$ and thus the new portfolio weights are: 
 \begin{equation}
     \mathbf{x}=\frac{1}{W}\begin{bmatrix}
         w_1 \\
         w_2 \\
         w_3 \\
         w_4
        \end{bmatrix}=\begin{bmatrix}
         0.224  \\
          0.224  \\
         0.299 \\
         0.254
        \end{bmatrix}
  \end{equation}
By means of historical simulation, we estimate the new VaR of the portfolio and then the new margin factor:
\begin{equation}
a_2^*=  \frac{0.0123}{0.2+0.0123}=0.26756
\end{equation}
Finally, we can re-evaluate the marginal availability at the beginning of day two considering the new amount invested and the new margining factor:
\begin{equation}
M_2^i=C_1^f-a_2^*W_2=  10329-0.26756 \times 33500= \$1365.8
\end{equation}
Notice that we have used the new value of the portfolio $C_1^f$. Since the new marginal availability is positive, we can allow the trade. 

Like we did before, we simulate the situation at the end of day two using historical returns.

\begin{figure}[H]
\center
\includegraphics[scale=0.8]{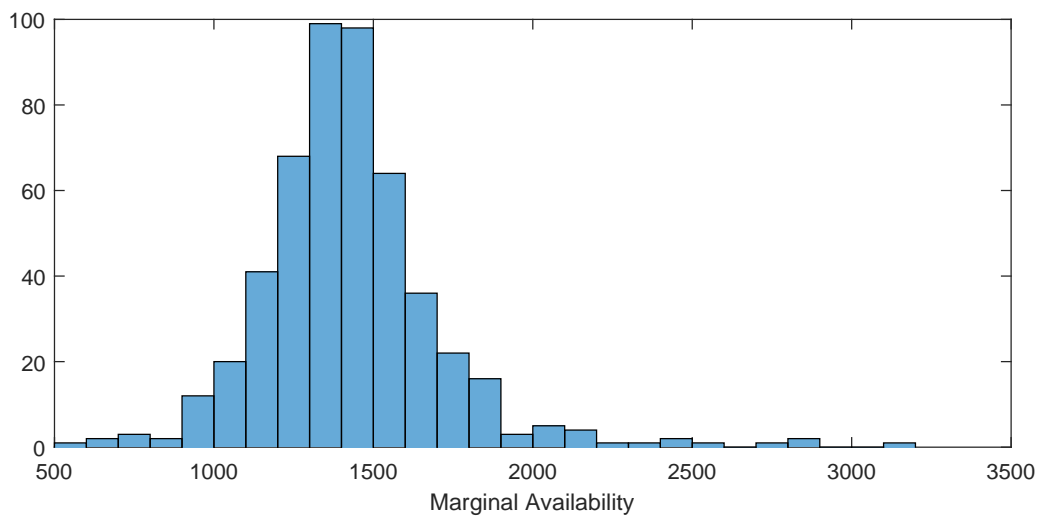}
\caption{Marginal availability distribution at the end of day two.}
\label{fig03}
\end{figure}

\begin{figure}[H]
\center
\includegraphics[scale=0.8]{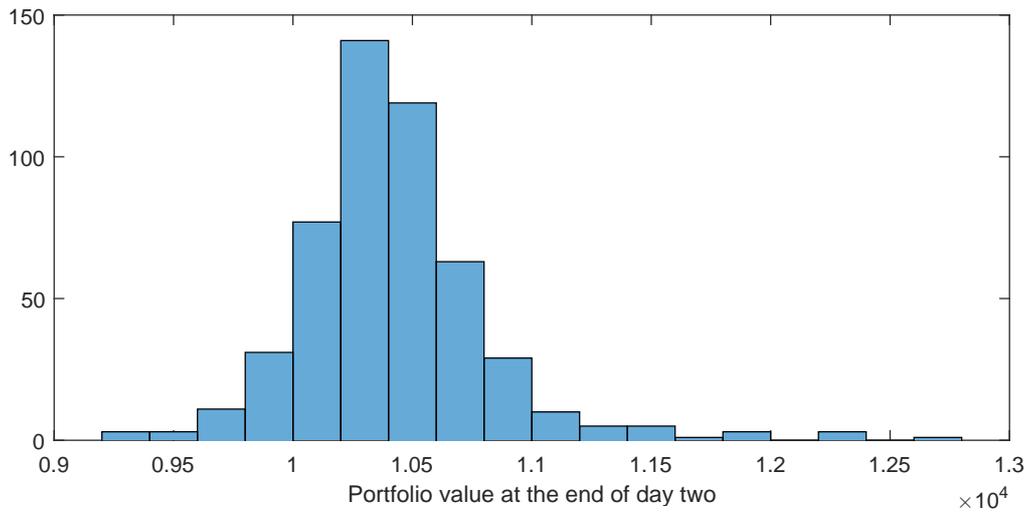}
\caption{Portfolio final value distribution at the end of day two.}
\label{fig03}
\end{figure}

Like we observed in the previous example, adding a protective put position has decreased the riskiness considerably and lowered the marining factor accordingly.

Of course, it could also happen that the marginal availablity becomes negative after an adverse price movement. In that case the client (or the firm) should close some positions or deposit more money in order to restore a positive margin.

\end{exmp}

\end{document}